# Single-walled TiO$_2$ Nanotubes: Enhanced Carrier-transport Properties by TiCl$_4$–Treatment

Imgon Hwang,[a] Seulgi So,[a] Mohamed Mokhtar,[b] Abdelmohsen Alshehri,[b] Shaeel A. Al-Thabaiti,[b] Anca Mazare,[a] and Patrik Schmuki*[a],[b]

[a] Department of Materials Science Engineering, WW4-LKO, University of Erlangen-Nuremberg, Martensstrasse 7, D-91058 Erlangen, Germany.

[b] Department of Chemistry, King Abdulaziz University, Jeddah (Saudi Arabia)

*Author to whom correspondence should be addressed.
E-mail: schmuki@ww.uni-erlangen.de
Tel: +49-9131-852-7575, Fax: +49-9131-852-7582

**Abstract**

In the present work we report significant enhancement of the photoelectrochemical properties of self-organized $TiO_2$ nanotubes by a combined "de-coring" of classic nanotubes followed by an appropiate $TiCl_4$-treatment. We show that, except for the expected particle decoration, a key effect of the $TiCl_4$ treatment is that the electron transport characteristics in $TiO_2$ nanotubes can be drastically improved, e.g. we observe an enhancement of up to 70 % in electron transport times.

*Keywords:* Photoelectric property, Photoanode, $TiO_2$ nanotube, Single-walled, Anodization

# 1. Introduction

Over the past decade, anodic $TiO_2$ nanotube layers have been widely explored in view of science and technology [1,2]. The main reason for such a rapid expansion of research is based on the easy formation of these highly ordered oxide nanostructures by a self-organizing electrochemical process. This coupled with the promising functional properties of the aligned oxide structure (optical, electronic, photoelectric, etc.) [1,3-5] paves the way to their application in dye-sensitized solar cells [6-8], photocatalysis [9,10], gas sensors [11,12], or biomedical devices [13,14].

Photoelectrochemical applications (solar cells, photocatalysis) are specifically derived from the semiconductive nature of $TiO_2$, i.e. the ability to form light induced mobile electron-hole pairs, in the material when subjected to super-band gap [1] illumination. As $TiO_2$ nanotubes can be easily varied through anodization parameters and annealing, there are a considerable number of reports that describe the influence of nanotube geometry and crystal structure on their photoelectrochemical properties [15-20]. Therefore, a large number of photoelectrochemical data has been obtained but, in general, only comparably low charge carrier mobilities have been reported for conventional anodic $TiO_2$ nanotubes [21].

Nevertheless, for most of these studies, typical fluoride-containing ethylene glycol based electrolytes were used for the growth of $TiO_2$ nanotubes and this leads, due to the nature of the anodizing process, to a double-walled structure of the nanotubes schematically shown in Figure 1(left panel) [22, 23]. The inner layer that is the layer formed at the oxide/electrolyte interface is less defined and considerably carbon-contaminated [24] due to the decomposition (and incorporation) of organic electrolytes during anodization at high voltages [23,24].

Only recently an approach was developed that can de-core such 'double-walled' tubes to a 'single-walled' morphology. The process is based on a selective etching process; i.e. the carbon rich inner shell is selectively etched out leaving the outer shell behind [25].

In the present work, we investigate the photoelectrochemical properties of these single-walled nanotubes in comparison to double-walled morphologies and explore the effect of a $TiCl_4$ treatment [26] on the photoelectrochemical properties of the tubes. We show that the treatment not only leads to an expected nanoparticle decoration but remarkably also significantly improves the electronic transport properties of the tubes. Thus $TiCl_4$ treatments on single-walled tubes can strongly improve the photoelectrochemical activity of $TiO_2$ nanotubes.

## 2. Experimental

Titanium foils (0.125 mm thickness, 99.6 % purity, Advent) were chemically polished by a solution containing hydrofluoric acid (HF, 40 %, Sigma-Aldrich), nitric acid ($HNO_3$, 70%, Sigma-Aldrich) and deionized (DI) water (1:4:2 vol. %), followed by washing with DI water and drying in nitrogen stream. Before measurements the polished Ti foils were sonicated in acetone, DI water and ethanol for 15 min, respectively, and then were dried in a nitrogen stream. The electrolyte consisted of 1.5 M lactic acid (LA, 90 %, Sigma-Aldrich), 0.1 M ammonium fluoride ($NH_4F$, 98 %, Sigma-Aldrich) and 5 wt. % of DI water in ethylene glycol (EG, 99.5 %, Sigma-Aldrich), as typically used for the fast growth of highly ordered nanotubular layers [27]. The first anodization was carried out in a two-electrode system with platinum foil as cathode and Ti foil as an anode, using 120 V from a high-voltage potentiostat (Jaissle IMP 88 PC) for an anodization time of 10 min. After anodization, samples were sonicated to remove the nanotubular layer and a dimpled structure was left on the Ti substrate, leading to more ordered $TiO_2$ nanotubes in the subsequent second anodization. The second anodization was carried out under the same conditions (electrolyte, voltage); however, for obtaining a 2.5 µm thick layer, the anodization time was 1 min at 60 °C – heating was performed at the backside of the Ti substrate using a temperature-controlled thermostat (HAAKE F3). After anodization, samples were washed in methanol and dried in nitrogen stream.

Single-walled $TiO_2$ nanotubes can be obtained from as-grown $TiO_2$ nanotubes by following procedure: The as-grown samples obtained after anodization are annealed at 150 °C for 1 h in air, with a heating and cooling rate of 30 °C/min using a Rapid Thermal Annealer (Jipelec JetFirst100). Afterwards, the annealed nanotubes are etched by piranha solution (1:3 vol. % of $H_2O_2$:$H_2SO_4$) at 70 ˚C for 330 sec. Samples were subsequently annealed at 450 ˚C for 1h (for annealing experiments also other temperatures were used, such as 350 and 600 ˚C). On the other hand, to directly obtain annealed double-walled $TiO_2$ nanotubes, as-grown nanotubes were annealed at 450 ˚C for 1 h. For the $TiCl_4$-treatment, single and double-walled $TiO_2$ nanotubes were immersed once or twice in 0.1 M $TiCl_4$ solution at 70 ˚C for 1 h. All treated samples were washed with DI water and ethanol, and then dried in nitrogen stream. After the $TiCl_4$ treatment, samples were annealed again, this second annealing step was initially performed at 450 ˚C for 10 min and later other temperatures were also investigated (350 and 600 ˚C).

A field-emission scanning electron microscope (FE-SEM, Hitachi SEM FE 4800) and a transmission electron microscope (TEM, Philips CM30 TEM/STEM) were used to perform the morphological characterization of the nanotubular samples. The structure and crystallinity of prepared samples were investigated by x-ray diffraction analysis (XRD, X'pert Philips PMD with a Panalytical X'celerator detector) using graphite-monochromized CuKα radiation ($\lambda = 1.54056$ Å). In order to confirm the amount of the carbon contents both single – and double-walled $TiO_2$ nanotubes, an X-ray photoelectron spectrometer (XPS, PHI 5600 XPS spectrometer) and an energy dispersive X-ray spectroscopy (EDX, fitted to SEM chamber) were used.

Photoelectrochemical characterization was performed with a 150 W Xe arc lamp (LOT-Oriel Instruments) with an Oriel Cornerstone 7400 1/8 m monochromator in both pure 0.1 M $Na_2SO_4$ and in 2 M MeOH containing 0.1 M $Na_2SO_4$ solution, at 500 mV in a three-electrode system with $TiO_2$ nanotubes as photoanode, saturated calomel electrode and platinum foil as a reference electrode and a counter electrode, respectively.

The range of photocurrent spectra were recorded from 300 to 600 nm.

Intensity modulated photocurrent spectroscopy (IMPS) measurements were carried out using a Zahner IM6 (Zahner Elektrik, Kronach, Germany) with an UV modulated light ($\lambda$ = 365 nm). The photoelectrochemical performance of the samples was analyzed in pure 0.1 M $Na_2SO_4$ solution, and in 2 M MeOH containing 0.1 M $Na_2SO_4$ solution in a three-electrode configuration, consisting of $TiO_2$ nanotubes as a photoanode, platinum wire as a counter electrode and a Ag/AgCl (3M NaCl) electrode as a reference electrode. The range of frequency was modulated from 1 mHz to 1 kHz.

## 3. Results and discussion

SEM and TEM images of typical TiO$_2$ nanotubes are shown in Figure 1(b) and 1(d), with a double-shell structure composed of an inner and an outer layer (as indicated in the top-view SEM image). The inner shell can be selectively etched, as illustrated in Figure 1(a), resulting in 'single-walled' TiO$_2$ nanotubes (see Figure 1(c) and 1(e)). The cross sections (insets in Figure 1) show that the removal process is successful over the entire nanotubular layer (≈2.5 μm), and both XPS (1(j)) and EDX (1(k)) results confirmed that most of the carbon contents was removed by an etching process. (The remnant amounts are within the range of natural contamination for both detection techniques) The double-walled nanotubes (DT) have an inner diameter of 91 nm at the top and 56 nm at the bottom, whereas for single-walled (ST) the inner diameter is approximately 105 nm over the entire tube length.

Both ST and DT were in some experiments treated with TiCl$_4$, on the one hand to replace the contamination layer by a TiO$_2$ nanoparticle layer (providing a high surface area hierarchical structure), and also to explore potential electronic effects of the TiCl$_4$ treatment.

The morphologies of the tubes after one-time TiCl$_4$ treatment (DT1, ST1) and after two-times TiCl$_4$ treatments (DT2, ST2) are shown in Figure 1(f) to 1(i) (all these structures are shown after annealing at 450 ˚C). Clearly for DT the TiO$_2$ nanoparticle decoration is less defined than for ST. This is even more apparent from DT2 that shows an increased nanoparticle decoration onto the tube surface and considerable morphological changes. Namely, when DT2 is compared to DT1 or DT (Figure 1(h) to 1(f) and 1(b)), it is evident that two-times TiCl$_4$ treatment leads to an overfilling of the DT-nanotube with nanoparticles. In contrast, for the ST, a defined nanoparticle loading is obtained for one-time and two-times TiCl$_4$ treatment with a clear nanoparticle layer on both the inside and outside of the tube wall.

In order to evaluate the photoelectrochemical properties, the photocurrent spectra (Figure 2(a) and 2(c)) and electron transport times (Figure 2(b) and 2(d)) were acquired in a classic aqueous electrolyte (Na$_2$SO$_4$) as well as in methanol containing electrolyte. Here the MeOH acts as a hole scavenger in order to reduce the influence of internal recombination (the fast transfer of holes to the MeOH electrolyte reduces their recombination probability with valence band electrons).

Figure 2(a) and 2(c) show the results as ICPE vs. wavelength curves, and Figure 2(b) and 2(d) show the intensity modulated photocurrent spectra (IMPS), for the bare nanotubular structures (DT0 and ST0) and for the TiCl4 treated ones (one-time decorated: DT1, ST1; two-times decorated: DT2, ST2). From these results it is apparent that irrespective of the measuring solution, single-walled nanotubes and double-walled nanotubes in their "native" state provide similar photocurrent magnitude but the ST provide faster electron transport kinetics (even more pronounced in the hole capture electrolyte). Most striking is, however, the difference if a single TiCl$_4$ treatment is applied, the ST show 65% increased efficiency compared to DT and a one decade faster electron transport than in the untreated state. The TiCl$_4$ treatment also increases the photoelectrochemical properties of the DT but only to a significantly lesser extent. The results thus show that not a higher surface area (a higher nanoparticle loading) is most responsible for the improved photoelectrochemical properties, but more importantly the TiCl$_4$ treatment provides an additional beneficial effect on the electron transport properties. The effect can clearly be observed in both pure 0.1 M Na$_2$SO$_4$ (Figure 2(b)) and MeOH addition electrolytes (Figure 2(d)).

Further we investigated the effect of annealing temperatures while keeping a fixed nanoparticle loading. (The annealing temperature can change the structure and the morphology of nanotubes, and thus considerably influences the photoelectric properties [28].) As mentioned before, nanoparticle decorated ST were exposed to a two-step annealing processes, first after de-coring and second after the $TiCl_4$ treatment. The goal of the second annealing step was to anneal the nanoparticles [29]. We investigated annealing temperatures of 350, 450 or 600 ˚C for 1h (first step) and 10 min (second step), and their influence on structure, morphology and photoelectrochemical performance. Figure 3 shows the X-ray diffraction patterns of the different two-step annealing treatments. When both annealing steps are performed at 450 ˚C, $TiO_2$ nanotubes are composed of anatase and a low amount of rutile; whereas annealing below 450 ˚C leads to a similar crystallinity for all nanotubes. Generally, rutile is detected only when annealing at temperatures over 450 ˚C.

The influence of the two-step annealing process on the morphology of nanoparticle decorated nanotubes can be seen in Figure 4, both in the cross-section and corresponding top-views (insets). $TiO_2$ nanotubes annealed at temperatures of 350 ˚C or 450 ˚C in both steps (Figure 4(a)-(d)) show a similar morphology. However, when using 600 ˚C in either or both of the annealing steps, the structure and morphology of nanotubes is considerably altered (besides the change in crystalline structure clearly evident in Figure 3). As a matter of fact, the formation of rutile starts from the bottom of the nanotubes [22, 30] and is accompanied by a sintering and collapse of the structures. Consequently, the actual thickness of remaining nanotubes is smaller. Furthermore, when using 600 ˚C in both steps (see Figure 4(g)), the nanoparticles cannot be distinguished as separate units anymore, meaning that they also were sintered into the tube wall.

The influence of the annealing treatments on the photoelectrochemical properties of one-time nanoparticle decorated single-walled $TiO_2$ nanotubes is shown in Figure 5a and 5b. ICPE results show the highest photocurrents for the ST nanotubes with both annealing steps performed at 450 ˚C (450˚C_1h_450˚C_10min sample). In line with this, IMPS data shows the fastest transport times for this annealing sequence.

Best photocurrent improvements were obtained if tubes and $TiCl_4$ decoration were annealed at 450 ˚C in air, providing a strong beneficial effect on the electric transport effect times. For higher temperature annealing, an established detrimental effect of conversion to rutile is observed. That is, the less rutile in the structure, the higher the photocurrent response of the sample. This can be ascribed to a considerably shorter carrier life time in rutile which reduces the number of charge carriers reaching the back electrode; consequently, a decrease of the photocurrent efficiency can be observed [31].

As a conclusion, the present results on one hand underline the detrimental influence of the inner tube shell in the $TiO_2$ nanotubes. By removing this shell and replacing it with "clean" $TiO_2$ nanoparticles, properties of $TiO_2$ nanotubes can be significantly increased. Most interestingly, however, is that the $TiCl_4$ treatment not only provides "clean" particle decoration but also leads to a considerable improvement of the electronic transport properties in the $TiO_2$ nanotube walls.

## 4. Conclusion

The photoelectrochemical properties of single-walled $TiO_2$ nanotubes (obtained by anodization and subsequent etching) with and without $TiCl_4$ treatment were compared to classic nanotubes. From photocurrent spectra and intensity modulated photocurrent spectroscopy (IMPS) it is evident that the single-walled nanotubes provide significantly higher IPCE values and shorter electron transport times than double-walled nanotubes. Moreover, it is observed that a $TiCl_4$ treatment of the single-walled $TiO_2$ nanotubes considerably enhances the photoelectrochemical properties. This enhancement could be explained not only by an enlarged surface area but is due mainly to beneficial electronic properties. This may be ascribed to e.g. surface state passivating effects using $TiCl_4$. These results suggest that coring followed by an optimal $TiCl_4$ treatment can significantly improve the performance of the tubes in virtually all photoelectrochemical applications, such as photocatalysis and dye-sensitized solar cells.

**Acknowledgement**

The authors would like to acknowledge ERC, DFG and the Erlangen DFG cluster of excellence for financial support, and this project was funded by the Deanship of Scientific Research (DSR), King Abdulaziz University, under grant no. 16-130-36-HiCi. The authors, therefore, acknowledge with thanks DSR technical and financial support.

**REFERENCES**

[1] K. Lee, A. Mazare, P. Schmuki, *Chem. Rev.* **2014**, *114*, 9385.

[2] X. Zhou, N. T. Nguyen, S. Ozkan, P. Schmuki, *Electrochem. Commun.* **2014**, *46,* 157.

[3] P. Roy, S. Berger, P. Schmuki, *Angew. Chem. Int. Ed.* **2011**, *50*, 2904.

[4] A. Fujishima, X. Zhang, D. A. Tryk, *Surf. Sci. Rep.* **2008**, *63*, 515.

[5] X. Chen, C. Li, M, Grätzel, R. Kostecki, S. S. Mao, *Chem. Soc. Rev.* **2012**, *41*, 7909.

[6] W. Guo, X. Xue, S. Wang, C. Lin, Z. L. Wang, *Nano Lett.* **2012**, *12*, 2520.

[7] J. R. Jennings, A. Ghicov, L. M. Peter, P. Schmuki, A. B. Walker, *J. Am. Chem. Soc.* **2008**, *130*, 13364.

[8] P. Roy, D. Kim, K. Lee, E. Spiecker, P. Schmuki, *Nanoscale* 2010, *2*, 45.

[9] M. Kalbacova, J. M. Macak, F. Schmidt-Stein, C. T. Mierke, P. Schmuki, *Phys. Status Solodi RRL* **2008**, *2*, 194.

[10] I. Paramasivam, N. Liu, H. Jha, P. Schmuki, *Small* **2012**, *20*, 3073.

[11] D. Kim, A. Ghicov, P. Schmuki, *Electrochem. Commun.* **2008**, *10*, 1835.

[12] S. Joo, I. Muto, N. Hara, *J. Electrochem. Soc.* **2010**, *157*, J221.

[13] J. Park, S. Bauer, K. von der Mark, P. Schmuki, *Nano Lett.* **2007**, *7*, 1686.

[14] S. Bauer, P. Schmuki, K. von der Mark, J. Park, *Prog. Mat. Sci,* 2013, *58*, 261.

[15] B. Liu, K. Nakata, S. Liu, M. Sakai, T. Ochiai, T. Murakami, K. Takagi, A. Fujishima, *J. Phys. Chem. C* **2012**, *116*, 7471.

[16] Y. Liu, J. Li, B. Zhou, J. Bai, Q. Zheng, J. Zhang, W. Cai, *Environ. Chem. Lett.* **2009**, *7*, 363.

[17] N. Mir, K. Lee, I. Paramasivam, P. Schmuki, *Chem.-Eur. J.* **2012**, *29*, 11862.

[18] S. P. Albu, D. Kim, P. Schmuki, *Angew. Chem., Int. Ed.* **2008**, *47*, 1916.

[19] D. Kim, A. Ghicov, S. P. Albu, P. Schmuki, *J. Am. Chem. Soc.* **2008**, *130*, 16454.

[20] J. M. Macak, H. Tsuchiya, A. Chicov, K. Yasuda, R. Hahn, S. Bauer, P. Schmuki. *Curr. Opin. Solid State Mater. Sci.* **2007**, *11*, 3.

[21] S. So, K. Lee, P. Schmuki, *Chem. Eur. J.* **2013**, *19*, 2966.

[22] S. P. Albu, A. Ghicov, S. Aldabergenova, P. Drechsel, D. LeClere, G. E. Thompson, J. M. Macak, P. Schmuki, *Adv. Mater.* **2008**, *20*, 4135.

[23] N. Liu, H. Mirabolghasemi, L. Lee, S. P. Albu, A. Tighineanu, M. Altomare, P. Schmuki, *Faraday Discuss.* **2013**, *164*, 107.

[24] H. Mirabolghasemi, N. Liu, K. Lee, P. Schmuki, *Chem. Comumm.* **2013**, *49*, 2067.

[25] S. So, I, Hwang, P. Schmuki, *Energy Environ. Sci.* **2015**, DOI: 10.1039/c4ee03729d.

[26] M. K. Nazeeruddin, A. Kay, I. Rodicio, R. Humphry-Baker, E. Muller, P. Liska, N. Vlachopoulos, M. Gratzel, *J. Am. Chem. Soc.* **1993**, *115*, 6382-6390.

[27] S. So, K. Lee, P. Schmuki, *J. Am. Chem. Soc.* **2012**, *134*, 11316.

[28] R. P. Lynch, A. Ghicov, P. Schmuki, *J. Electrochem. Soc.* **2010**, *157*, G76.

[29] K. Zhu, N. R. Neale, A. F. Halverson. J. Kim, A. J. Frank, *J. Phys. Chem. C* **2010**, *114*, 13433.

[30] A. Mazare, I. Paramasivam, K. Lee, P. Schmuki, *Electrochem. Commun.* **2011**, *13*, 1030.

[31] P. Roy, D. Kim, I. Paramasivam, P. Schmuki, *Electrochem. Commun.* **2009**, *11*, 1001.

**Figure captions**

**Figure 1.** (a) Schematic illustration of single-walled TiO$_2$ nanotube formation by etching process. SEM images for (b) single - and (c) double-walled and TEM images for (d) single – and (e) double-walled TiO$_2$ nanotubes. The inset shows a SEM cross section of the double and single-walled TiO$_2$ nanotubes, respectively. SEM images for (f) once and (g) twice TiCl$_4$ treated double-walled, (h) once and (i) twice TiCl$_4$ treated single-walled morphology of TiO$_2$ nanotubes (top view along the length). Carbon amount of single – and double-walled TiO$_2$ nanotubes by (J) XPS and (K) SEM-EDX.

**Figure 2**. (a) IPCE and (b) IMPS for different nanoparticle loading in single and double walled TiO$_2$ nanotubes in pure 0.1 M Na$_2$SO$_4$; (c) IPCE and (d) IMPS for the same nanotubular structures in 0.1 M Na$_2$SO$_4$ with 2 M MeOH addition. The exact treatment of TiCl$_4$ cycles is: without (T0), with once (T1) and twice (T2), respectively.

**Figure 3**. X-ray diffraction patterns of 0.1 M TiCl$_4$-treated single-walled TiO$_2$ nanotubes annealed in the first step for 1 h (350, 450, and 600 ˚C) and in the second step (350, 450, and 600 ˚C), respectively, for 10 min in air environment: A, anatase; R, Rutile; T; Ti substrate.

**Figure 4**. SEM cross section and top view (insets) images of one-time TiCl$_4$ treated

single-walled TiO$_2$ nanotubes annealed at (a) 350 ˚C for 1 h and 350 ˚C for 10 min, (b) 350 ˚C for 1 h and 450 ˚C for 10 min, (c) 450 ˚C for 1 h and 350 ˚C for 10 min, (d) 450 ˚C for 1 h and 450 ˚C for 10 min, (e) 450 ˚C for 1 h and 600 ˚C for 10 min, (f) 600 ˚C for 1 h and 450 ˚C for 10 min, and (g) 600 ˚C for 1 h and 600 ˚C for 10 min.

**Figure 5.** Results of (a) IPCE and (b) IMPS in 0.1 M Na$_2$SO$_4$ with 2 M MeOH addition, for one-time decorated single-walled TiO$_2$ nanotubes annealed in two steps: first step for 1h (350, 450, or 600 ˚C) and second step for 10min (350, 450, or 600 ˚C), in an air environment.

**Figure 1.**

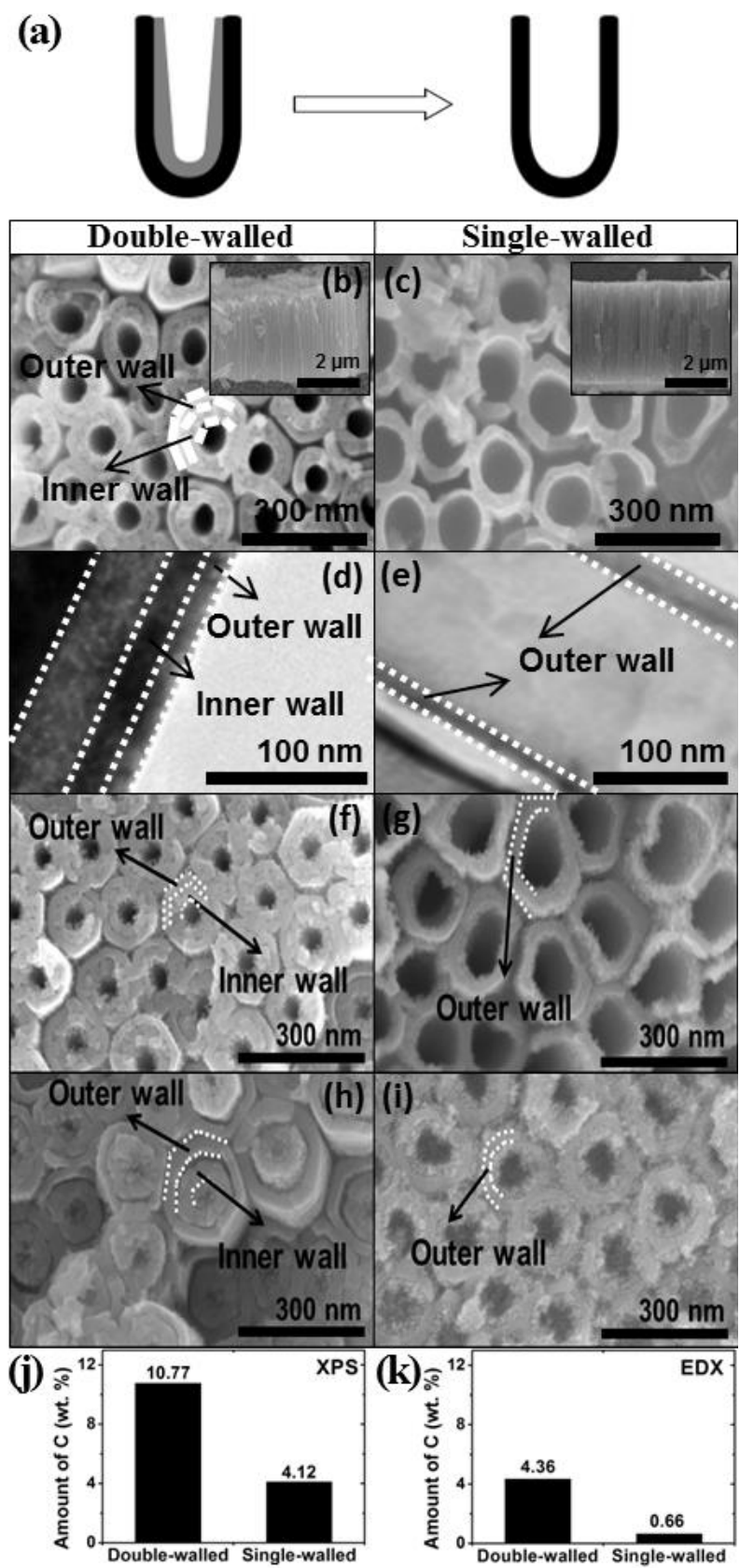

**Figure 2.**

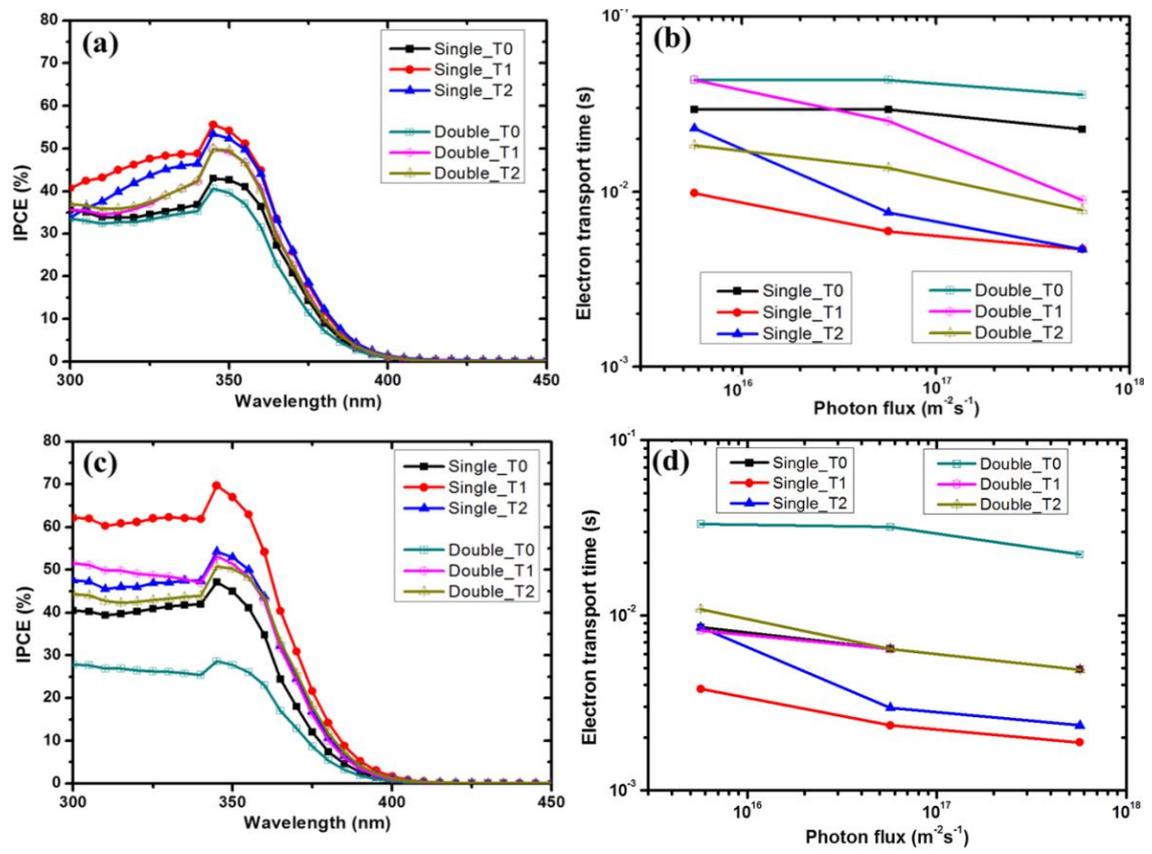

**Figure 3.**

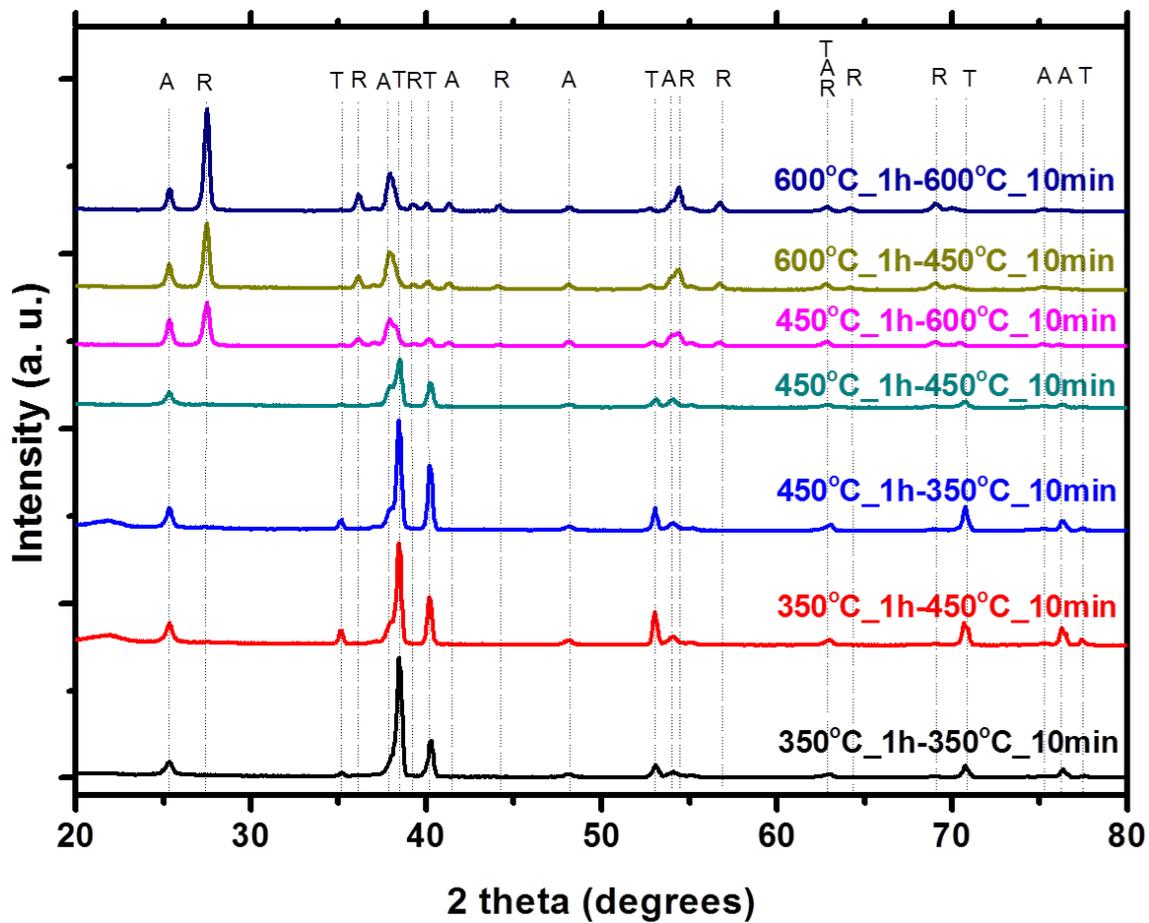

**Figure 4.**

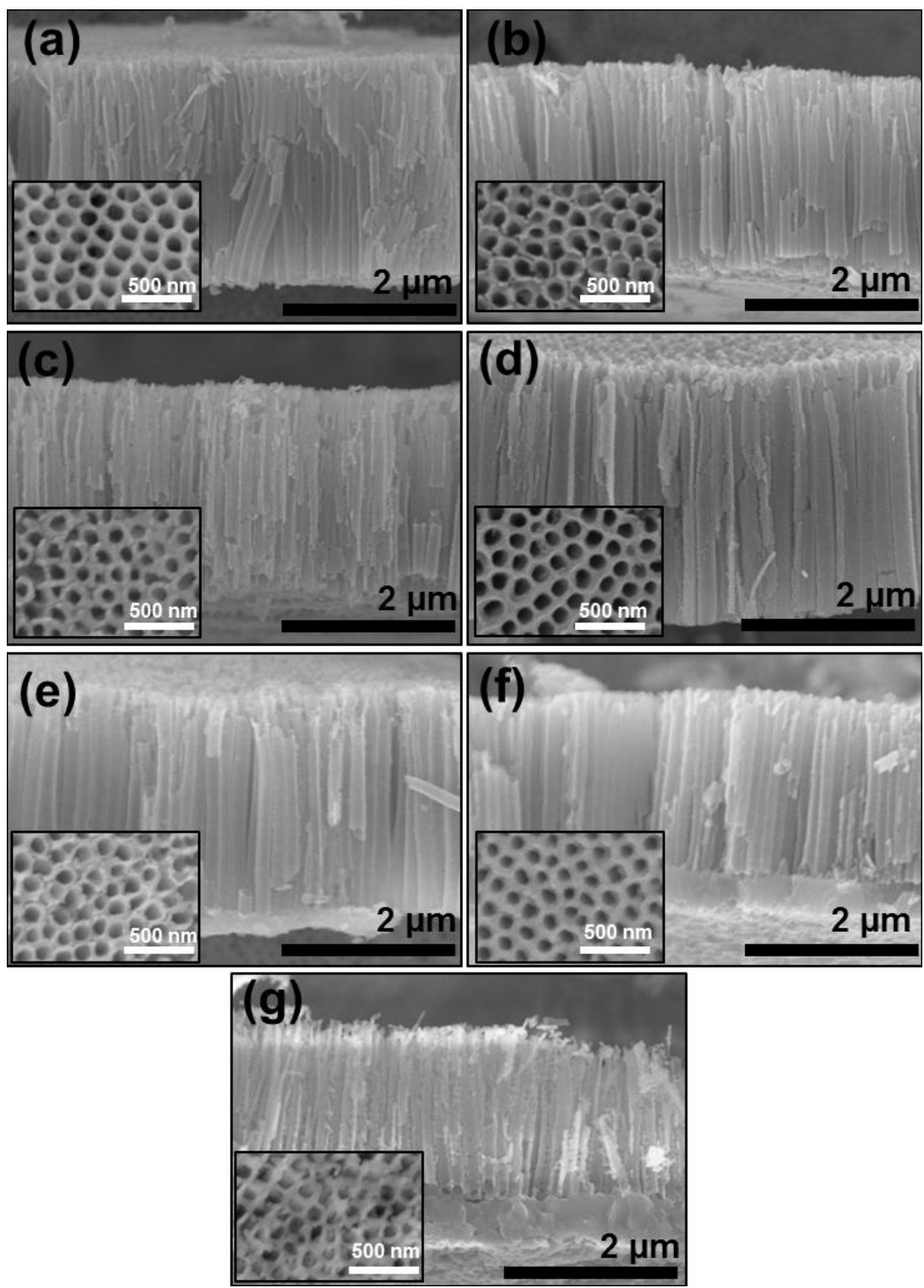

**Figure 5.**

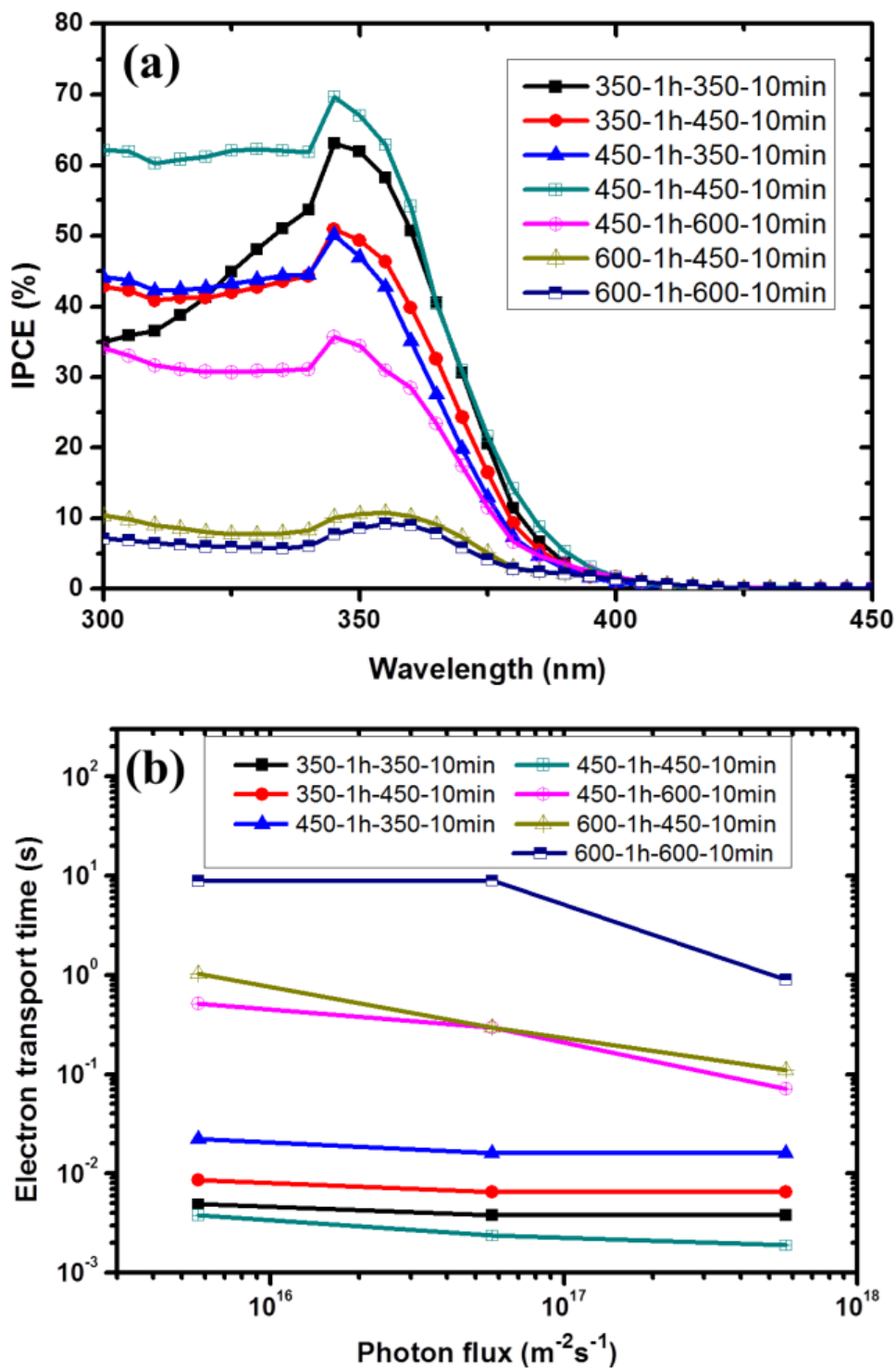